# Adapting to Educate: Conversational AI's Role in Mathematics Education Across Different Educational Contexts

Alex Liu
University of Washington
alexliux@uw.edu

Lief Esbenshade
University of Washington
lief@uw.edu

Min Sun
University of Washington
misun@uw.edu

Shawon Sarkar
University of Washington
ss288@uw.edu

Jian He
Hensun Innovation LLC
kevin@hensuninnovation.com

Victor Tian
University of Washington
ztian27@uw.edu

Zachary Zhang
Hensun Innovation LLC
zac@hensuninnovation.com

## ABSTRACT

As educational settings increasingly integrate artificial intelligence (AI), understanding how AI tools identify – and adapt their responses to – varied educational contexts becomes paramount. This study examines conversational AI's effectiveness in supporting K-12 mathematics education across various educational contexts. Through qualitative content analysis we identify educational contexts and key instructional needs present in educator prompts and assess AI's responsiveness. Our findings indicate that educators focus their AI conversations on assessment methods, how to set the cognitive demand level of their instruction, and strategies for making meaningful real-world connections. However, educators' conversations with AI about instructional practices do vary across revealed educational contexts; they shift their emphasis to tailored, rigorous content that addresses their students' unique needs. Educators often seek actionable guidance from AI and reject responses that do not align with their inquiries. While AI can provide accurate, relevant, and useful information when educational contexts or instructional practices are specified in conversation queries, its ability to consistently adapt responses along these evaluation dimensions varies across different educational settings. Significant work remains to realize the response differentiating potential of conversational AI tools in complex educational use cases. This research contributes insights into developing AI tools that are responsive, proactive, and anticipatory, adapting to evolving educational needs before they are explicitly stated, and provides actionable recommendations for both developers and educators to enhance AI integration in educational practices.

## 1. INTRODUCTION

K-12 mathematics education is pivotal in cultivating the cognitive and problem-solving skills essential for students' success in high er education and their future career [16, 38]. Studies underscore the transformative potential of robust mathematics instruction on student achievement [36]. Recent declines in student performance reported by the National Assessment of Educational Progress (NAEP), along with varying trends among achievement levels, underscore the urgent need for individualized and context-specific instructional support [32]. Mathematics educators face ongoing challenges designing lessons and tasks that cater to all student needs while adhering to rigorous educational standards [14]. Effective lesson preparation, which enables educators to translate complex mathematical concepts into pedagogically sound and actionable classroom strategies, is critical [21]. However, the development of high-quality lessons is often hampered by significant constraints, such as time, resources, and a lack of established metrics for evaluating lesson quality [19, 22, 39].

In response to these challenges, Artificial Intelligence (AI) built on Large Language Models (LLMs) has emerged as a potent tool, with the potential to transform how educators develop and refine their professional practices [31]. This integration of AI within educational settings has facilitated unprecedented access to data on educators' preparatory activities, whether in lesson planning or active teacher learning. The wealth of data from educator-AI interactions offers insights into various pedagogical approaches, catalyzing the development of support tools that enhance instructional effectiveness. Contemporary research has increasingly focused on harnessing AI to develop tools that support dynamic instructional needs, including novel instructional modalities to provide real-time feedback (e.g., [15, 17, 50]), collaboration and group discussion facilitators (e.g., [25]), and assist assessment grading through learning analytics (e.g., [18, 30]).

Despite these technological advances, a noticeable gap persists in our understanding of how educators interact with AI tools during the lesson preparation phase. Existing research largely studies AI's real-time data analytics potential and the effect of feedback within the scope of classrooms (e.g., [13, 43]). However, there is a dearth of empirical studies that delve into the integration of AI within the broader spectrum of teacher' professional workflows. There has been considerable exploration of student-AI interactions within intelligent tutoring systems (e.g., [27, 37, 40]), but the dialogues between educators and AI as they refine lesson content and pedagogical strategies remain relatively underexplored. These interactions are rich data sources that record teacher learning opportunities, with AI acting not just as a tool, but also as a tutor or peer collaborator, aiding in the development of instructional strategies tailored to classroom needs [24]. The extent to which these AI platforms adapt to specific educational contexts and effectively support a range of instructional goals is still not fully understood, leaving uncertainties regarding AI's generated contents ability to meet contextualized educational demands and produce outcomes that are accurate, relevant, and useful [23].

As instruction-fine-tuned LLMs like ChatGPT and Anthropic Claude are promoted for educator use, there is an expectation by teachers that these technologies can adapt dynamically by generating responses based on specific educator commands [27]. The critical issue, however, lies in the models' ability to handle abstract and complex conceptual information accurately—a crucial aspect for mathematics educators who rely on the precision of content to foster understanding and minimize confusion [47]. The challenge is exacerbated by generative AI's potential inaccuracies, such as a misrepresenting numerical values (e.g., 3.11 > 3.9), which can lead to educational setbacks for student learners and potentially deter technology adoption. Additionally, mathematics educators are often tasked with addressing different learning needs, learning progression, and learning styles, while providing social emotional and behavioral support to cultivate a positive learning environment [19]. In particular, for educators teaching under-represented student groups the generic responses generated by LLMs, based on the most commonly available data, may not align with the specific instructional needs or contexts, thereby perpetuating uneven educational experiences [1, 34]. To truly harness the potential of AI to meet the specific, individual, needs of educators and their students, it is imperative to analyze the emerging corpus of educator-AI interaction data. This will allow us to discern patterns of mathematics educators' contextualized instructional practices and assess the adaptability of AI across various educational scenarios.

In this study, we leverage large-scale mining of educational conversation data over 3,400 thousand teacher-AI interactions and undertake an LLM based content analysis of educational contexts and instructional practices as they manifest within these interactions. By automating the assessment of AI response quality across both general and specific contexts and practices, we identify AI's adaptability of instructional support to align with the specific needs of educators shaped by their distinctive educational environments. We address following questions:

- RQ1. Conceptualization of AI in Educator Support: How do classroom settings and specific student needs influence educators' requirements for AI support within their instructional practices?
- RQ2. Measuring AI Response Quality: How can the quality of AI responses in educational dialogues be quantitatively assessed to determine their professional usability?
- RQ3. Adaptability of AI Responses across Educational Contexts: How effectively can AI responses adapt to educational contexts to support instructional practices as they emerge in conversation flows?

Our findings aim to provide insights for the usage and usability of AI tools in educational settings, which serve not only as supplemental professional development for educators' technological skills but also can be enhancers of pedagogical toolkits. These insights hold the potential to reduce the time educators spend on non-student-facing tasks, thereby increasing their capacity to engage more directly with students and enhancing overall instructional quality and job satisfaction [7]. Moreover, this research is crucial for educational technology developers to design responsive, proactive, and anticipatory tools that align with the dynamic needs of a comprehensive mathematics instruction landscape.

# 2. RELATED WORK

The integration of AI in educational settings has captured significant attention in recent years, reflecting a growing consensus on its transformative potential. This section reviews the existing literature on AI applications within education, with a focus on automated content analysis and labeling with LLMs, the assessment of AI response quality, and the adaptability of AI tools across diverse educational applications.

## 2.1 Automated Content Analysis and Labeling

Researchers have increasingly utilized text as a form of qualitative data to explore complex social phenomena [11]. The integration of LLMs into text analysis has expanded significantly, bolstered by their advanced reasoning and interpretative capabilities [29]. These LLM-assisted automated text analysis procedures have extended the reach of qualitative research, overcoming the limitations of traditional Natural Language Processing (NLP), such as topic modeling methods, which have been critiqued for producing low-quality codes or failing to capture nuanced insights that are evident to human analysts. Such methods have also struggled to articulate the reasoning behind their analytic outcomes [3]. Facilitated by LLMs, automated coding approaches now enable precise text analysis and labeling on a large scale, executed within relatively short time frames.

LLMs have been applied in inductive coding, constructing theoretical meaning from the text [2]. Inductive, bottom-up coding techniques grounds codes within the research context, allowing meanings to emerge from the codebook development process by identifying frequent, dominant, or significant themes [5]. In a recent study [3], researchers incorporated GPT-4 to evaluate the impact of human and LLM involvement in codebook development on the coding quality of 60-minute math tutoring session transcripts. Results indicated that using LLM for preliminary codebook development with human refinement required less time and generally outperformed other methods in terms of clarity of codes and mutual exclusivity. Other researchers [10], noted GPT-3.5-turbo's ability to infer key themes from a "teaching" dataset, revealing nuanced themes that were initially overlooked by analysts. Thus, combining LLM codebook development with manual inspections is particularly effective for studies involving large-scale and complex data. Integrating LLMs into this process necessitates meticulous consideration of the timing of LLM involvement, the design of LLM prompts, and the role of human researchers to ensure data analysis reliability and validity.

LLMs have been used for deductive coding, applying theoretical concepts to interpret text [2]. Liu and Sun [28] employed GPT-4 to code transcripts of 24 interviews with educational stakeholders, using a developed codebook of 28 codes nested within 8 broad educational aspects. Their findings demonstrated that compared to Latent Dirichlet Allocation (LDA), LLMs were more capable of understanding themes within the text and more closely aligned with human qualitative researchers' results, achieving a hit rate of 74% at the detailed code level and 96% at broad code levels. These results underscored LLMs' ability for consistent judgment and potential to supplement human interpretations. Their findings are consistent with other studies utilizing LLMs for deductive coding [44, 49], where LLM's consistent inference and reasoning, combined with appropriate human supervision, potentially enhance the construct validity and reliability of coding results [48].

In addition to qualitative analysis, studies (e.g., [9, 42]) have shown that LLMs analyze input based on criteria. A comparison of GPT-3.5's evaluation results with humans using the same criteria for the International English Language Testing System (IELTS) academic writing confirmed that LLMs can deliver reliable and criterion-based grading [9]. However, the quality of rubrics for evaluation is crucial. These studies highlight the importance of developing precise rubrics for LLM applications, key to achieving good inter-rater consistency with benchmarks and demonstrating excellent test-retest reliability, thereby systematically assessing the quality and relevance of content. The creation of clear and specific rubrics tailored for educational contexts is essential for enhancing the accuracy and effectiveness of automated content analysis, ensuring consistent and objective analysis, and providing a robust framework for interpreting complex educational data.

## 2.2 Measurement of AI Response Quality

The assessment of AI-generated responses in educational settings is critical to ensure effective instruction and support. We summarize studies to develop metrics to evaluate the accuracy, relevance, and usefulness of these responses. These quality measures are essential for AI responses support educators in meaningful way that maintains instructional integrity and enhances teaching effectiveness.

**Accuracy** is a fundamental metric across various fields but takes on special significance in educational contexts. After the release of commercialized LLMs, concerns have risen over hallucinations, where LLMs generate plausible yet factually incorrect or nonsensical information, as a significant danger, leading to hesitancy in adopting these technologies [46]. In general AI applications, accuracy is benchmarked against factual correctness and logical consistency. For mathematics educators' usage of AI, they require factual correctness regarding the pedagogical aspects in the generated content and the reasoning accuracy for subject knowledge and assessment gradings [47]. Specifically, in mathematics reasoning, the accuracy of AI responses must encompass both computational and conceptual correctness to support the integrity of instruction. Within educational settings, ensuring accuracy also means no misinformation contained in AI-generated content regarding established learning standards, curricula, and pedagogy frameworks. A critical factor in maintaining the educational value of AI-assisted teaching [41].

**Relevance** is a criterion in evaluating the quality AI-generated responses as how well and comprehensive they address the requests. For general use AI, relevance might simply pertain to user satisfaction or query resolution. However, for domain specific AI, the evaluation of relevance reflects how well AI handle domain specific inquiries. A study by Hamidi, which evaluated an AI chatbot's relevance in healthcare by measuring how relevant is the answer to the to specific patient inquiries regarding Electronic Health Records (EHRs) by ask and whether the response address the key medical concepts and details in the questions. In education, relevance demands a precise alignment of AI-generated responses with pedagogical objectives and the intended learning outcomes that educators aim to achieve, by matching teaching methods and assessment tasks with the specific instructional needs [4]. When educators interact with AI, they not only seek answers to content-specific questions of a pedagogy or a concept but also expect solutions that address their instructional challenges. Therefore, a relevant AI response must go beyond merely providing factual information. It must interpret the key concepts and details embedded in educator queries, understand the context of these educational needs, and tailor responses that support specific needs.

**Usefulness.** The concept of usefulness in AI-generated response within educational settings ensure that AI tools effectively support professional tasks, thus influencing both the user experience and adoption of these technologies [45]. Drawing on principles from the widely adopted System Usability Scale (SUS), developed by Brooke [6], it is critical for AI-powered educational tools to be perceived as "easy to use." This means that AI responses should be well-integrated with various functions within the educational field, practical, and not overly complex, reducing the needs for additional support outside the technological system from educators. For AI-powered instructional support, the efficacy of AI responses is achieved by reducing educators' cognitive load and providing step-by-step instructions streamlining lesson preparation and execution. This is analogous to AI tutors in mathematics that decompose complex problems into simpler steps, markedly improving student comprehension and learning outcomes. In addition, the usefulness of AI extends to facilitating direct applicability of generated content, including copy-editing functions like formatting responses in accordance with specific educator-defined requirements, such as tables and worksheets. The usefulness of AI in educational settings, therefore, depends on its ability to deliver practical, straightforward solutions customized to meet the distinct needs of users.

## 2.3 AI Use in for Instruction

AI's potential to transform educational practices through support in various instructional capacities has been increasingly recognized. For instance, Tissenbaum and Slotta [43] provide critical insights into how AI-powered tools can be integrated within a high school physics class. The study highlighted the effectiveness of AI-powered feedback systems in enhancing classroom management, boosting student engagement, and facilitating collaborative learning. Holstein et al [18] underscored the value of real-time feedback provided by AI-powered analytic tools in dynamic classroom settings, which equip teachers with actionable insights into student learning. Another, similar, study suggests that timely AI-generated classroom data-driven support, particularly combined with clear and consistent curriculum guidance, can enable teachers to make incremental adjustments that enhance their teaching effectiveness [24].

AI-based chatbots have been adapted and utilized in mathematics education. Nguyen et al [33] developed a chatbot that guides students through step-by-step problem-solving in high school mathematics, illustrating the broader application of AI in tutoring. Additionally, various adaptive learning tools are leveraging AI to create personalized learning experiences by tailoring content, feedback, and pacing to individual student needs [12, 20]. Dabingaya's study [8] on the effectiveness of AI-powered adaptive learning platforms in mathematics education focused on the positive reinforcement between student engagement and learning outcomes. While several other studies have compared adaptive learning tools that act as tutors to traditional teaching approaches with educators in classrooms, we have seen fewer studies for adapting AI tools to support the ongoing professional learning and practice needs of educators. One example for such conversational AI tools has been implemented in teacher preparation programs, offering preservice mathematics teachers practical, real-time teaching experience in scenarios designed to be authentic, meaningful, and open-ended. To ensure the chatbot responded appropriately, the researchers manually analyzed the training data, categorizing it into the smallest meaningful user intents and crafting corresponding responses for each intent [26]. Gaps remain in understanding how educators prepare for actual classroom instruction and whether current AI tools support authentic and meaningful interactions within conversations across a large sample of educators with different characteristics.

# 3. METHODOLOGY

## 3.1 Data Collection

We obtained 86,739 deidentified dialogues between educators and AI agents on a platform specifically designed for educator use that occurred between May 2024 and February 2025. To refine the dataset and select representative samples from this large-scale dataset, we used LLMs to identify specific subject areas within each conversation. We focused our analysis on mathematics instruction, and excluded multidisciplinary or interdisciplinary conversations, selecting only those in which mathematics was the sole subject discussed. We also manually verified the subject focus within the conversations in the sample. As a result, our sample comprises 3,429 conversations containing 11,173 requests and responses explicitly related to mathematics instruction, involving 939 K-12 mathematics educators from 47 states in the U.S..

We structured the dialogues within each conversation into sequential trios of educator requests, AI responses, and educator follow-up requests after AI responses. This structure allowed us to capture not only how AI addressed educators' needs but also provided insights into educators' immediate reactions to AI outputs. We coded each sequence with a set of labels. We identified the educational context and instructional practices queried in the educator requests. We assessed AI response quality across the dimensions of accuracy, relevance, and usefulness, and categorized the type of content offered by the AI, such as clarifying questions, actionable guidance, or information provision. Additionally, we recorded whether these responses were accepted or rejected by educators to gauge the responses' usability and applicability within specific educational contexts from educators' perspectives.

## 3.2 Content Analysis and Labeling

To analyze the interactions between educators and AI on this dedicated educational platform, we employed LLM for qualitative content analysis, including inductive coding and deductive coding approaches [1]. The automated method was chosen for its efficacy and time-efficiency in uncovering patterns and themes within the large sample of textual data, which is essential for understanding the nuanced educational contexts and instructional practices that emerge during these interactions [3, 28]. This analytical approach allows a deep understanding of how AI has been applied for educational practices and how they respond to the dynamic needs of educators.

### *3.2.1 Inductive Coding for Educational Contexts*

We conducted the inductive qualitative coding (Figure 1) by human-LLM collaboratively identifying the educational contexts, which included emerging student and teacher needs. Following the recommendation from previous work [3, 10], we initiated an open coding approach and tasked the LLM (Claude 3.5 Sonnet) with recognizing the educational needs revealed from a subset of conversations (N=500). We then manually reviewed the open-ended codes, sorting and classifying them into distinct groups. This phase of coding was specifically aimed at elucidating both student and classroom needs for AI support. Through this process, we compiled a list of identified educational contexts.

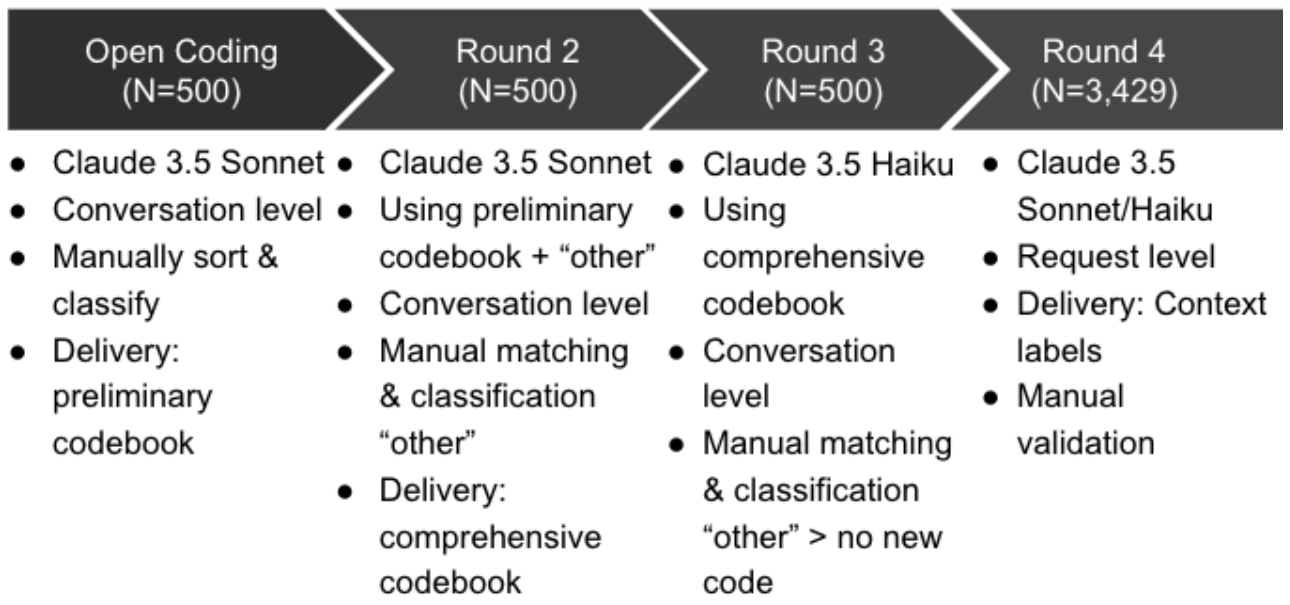

**Figure 1. Inductive Coding Workflow for Education Context**

Subsequently, we developed a preliminary codebook with code names and brief descriptions to facilitate the labeling of conversations. In the second round of coding, involving a sample size of a further 500 conversations, we allowed the LLM the flexibility to label unidentified contexts as "Other," providing brief specifications for any educational contexts it identified that were not listed in the codebook. Upon receiving the results, we manually matched the "Other" labels to existing codes if applicable, based on their descriptions and the content in original requests. For those "Other" labels without clear matches, we further sorted and classified them into new coding groups and incorporated these into the comprehensive codebook.

In the third round of coding using the comprehensive codebook, with a sample size of an additional 500 conversations, we also offered LLM (Claude 3.5 Haiku) the option of elaborating on "Other" contexts that were not in the codebook. After manually inspecting the resulting "Other," no new codes for student needs or classroom settings were added to the codebook, indicating a stabilization in the coding process. We then applied the comprehensive codebook, which included 11 educational contexts, to label the requests within all conversations from the entire dataset of 3,429 conversations, using the same LLMs (Claude 3.5 Sonnet and Claude 3.5 Haiku). By employing the same LLMs, we aimed to capture the educational contexts for user requests as they were revealed to the AI during the conversations. We aggregated the request-level results to compare them with the conversation-level outcomes. A 93% overlap between these levels demonstrated the feasibility of our automatic coding approach with our data structure.

### *3.2.2 Deductive Coding for Instructional Practices*

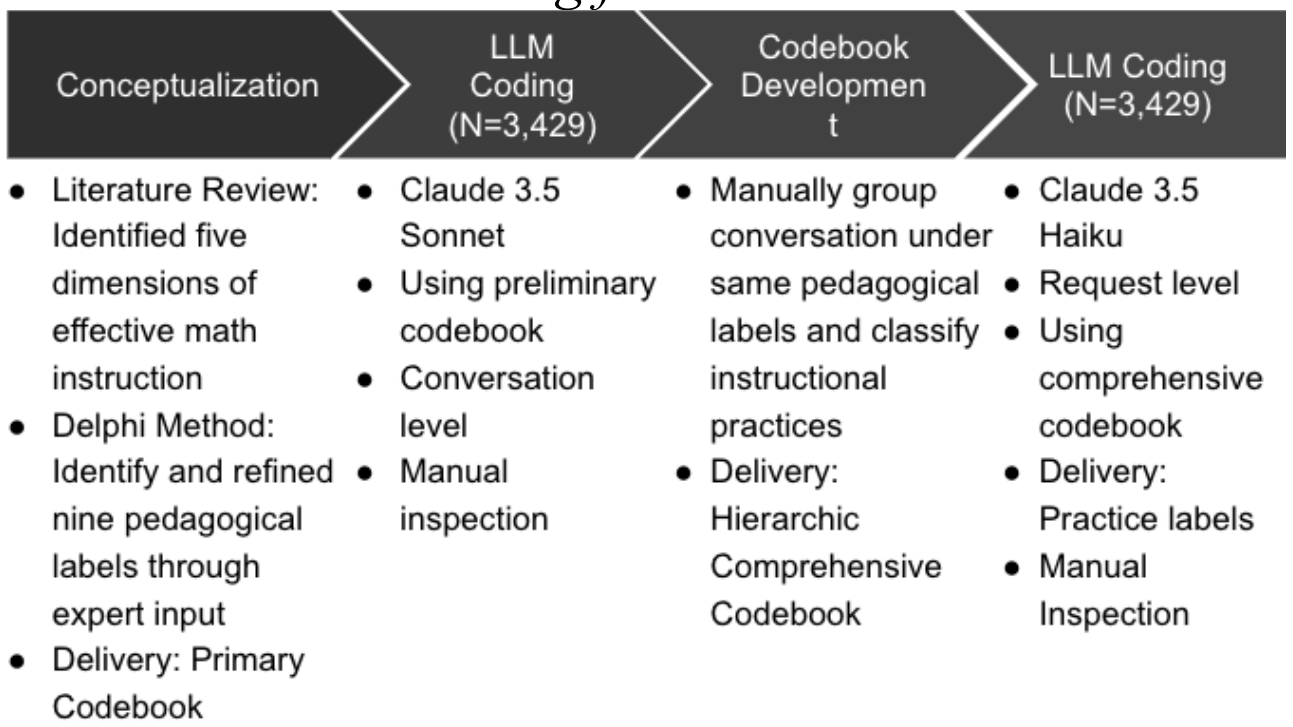

**Figure 2. Deductive Coding Workflow for Instruction Practices**

We adopted deductive coding (Figure 2) for recognizing effective instructional practices in the educator-AI conversations. To establish a structured and comprehensive codebook, we conducted extensive literature review to identify five dimensions of effective mathematics instruction from existing research. Then, utilizing the Delphi method, which included surveys and expert panel discussions with educational researchers and practitioners, we delineated nine pedagogical labels across these five dimensions [35]. These labels were then employed to annotate the pedagogical strategies that educators discussed with AI.

1) Content Rigor: (1) Focuses on the complexity of mathematical concepts and alignment with educational standards; (2) high cognitive demanding tasks.
2) Engagement: (1) Assesses the inclusion of student-centered learning opportunities and interactive elements; (2) meaningful connections to real-world examples or problems; (3) collaborative and group learning; (4) using multimedia and other innovative tools.
3) Differentiate Scaffolding: Evaluates how well the lesson plans accommodate all learners, such as students with disabilities and English language learners (ELLs).
4) Assessment Integration: (1) Examines the organization of the lesson, the clarity of instructional objectives, and the quality of supporting materials; (2) provide feedback to students and guide adjustment to future instructional plans.

Through this codebook development and LLM labeling process (Claude 3.5 Sonnet), we identified specific pedagogical aspects covered in the conversations. After grouping conversations around pedagogical themes, we conducted an analysis to pinpoint the specific instructional practices referenced in educator requests. Within each broader pedagogical category, we recognized several instructional practices commonly sought by educators, either for instructional implementation or lesson material generation.

As a result, we developed a hierarchical codebook of effective mathematics instruction, which consists of an upper layer of nine pedagogical aspects and a lower layer of 30 instructional practices. This codebook was utilized to relabel (Claude 3.5 Haiku) educator requests using the lower-layer instructional practices, according to the upper-layer pedagogy aspects identified within conversations. This structured approach allows for a nuanced understanding of the instructional dynamics within given pedagogical aspects and supports a targeted analysis of AI's role in facilitating educational practices.

#### *3.2.3 AI Response Evaluation*

We classified and evaluated AI responses within each educator-AI interaction based on their effectiveness, assessed by an LLM across three measures: accuracy, relevance, and usefulness. Each criteria-based measure was rated on a scale summing to a maximum of 3 points:

1) Response Accuracy: the Response must be factually and logically correct and free of errors or misinformation (3=Yes, 0=No)
2) Response Relevance: Fully addresses the original questions or instructional tasks (1=Yes, 0=No). Specific to the educational context (1=Yes, 0=No). Focused on the user's needs (1=Yes, 0=No)
3) Response Usefulness: Offers practical suggestions easily implemented in educational settings (1=Yes, 0=No). Reduces cognitive load with clear, step-by-step instructions (1=Yes, 0=No). Formats output as requested by users or consistent with their contexts (1=Yes, 0=No).

To enhance the evaluation of contextualized quality, the LLM was provided with corresponding requests when assessing the responses. This structured approach allows us to compile effectiveness scores for each AI response, providing a quantifiable measure of how well the AI's responses meet educators' needs. Further, to investigate the type of assistance provided, we categorized AI responses into the information, actionable guidance, and questions.

Additionally, we monitored the dynamics following each AI response, documenting instances where educators explicitly accepted the response by expressing gratitude or complimenting, or rejected it by correcting errors, pointing out mistakes, or refusing to answer questions proposed by the AI in their follow-up requests. This allowed us to assess the immediate adequacy of the response and test the validity of our AI response evaluation. Our hypothesis posits that AI responses explicitly rejected by educators should score lower on the evaluation measures compared to other responses, while explicitly accepted responses should score similarly or slightly higher compared to those not explicitly accepted or rejected, as educators may accept an AI response without overtly indicating their acceptance.

#### *3.2.4 Coding Validation*

For codebook develop in inductive coding, we employed a LLM complemented by a manual refinement to compile codebook for educational contexts. After obtaining comprehensive codebooks, for both inductive and deductive coding, we manually read through and verify randomly selected sample to validate LLMs results and made necessary corrections. Additionally, in our process, we have subsample of which conversations and requests were labeled using LLMs. This allowed us to take an extra step to validate coding results by comparing the aggregated request labels with conversation-level labels. Misalignments were manually examined and corrected, ensuring that the labels accurately reflected the content.

This dual-layer coding process not only facilitated rapid data processing but also ensured robust data interpretation, allowing for necessary adjustments based on researchers' educational domain knowledge. This method enhanced our capacity to accurately capture the complexities of AI interactions in educational settings within a limited time frame. As a result, we compiled a comprehensive database of categorized conversations, which included labels for educational contexts, pedagogical aspects, instructional practices, and types and evaluations of AI responses. This categorized data now forms the foundation for all subsequent analyses.

To validate quality measures and ensure that our metrics accurately reflect educators' perspectives on AI response quality and applicability, we specifically marked AI responses that were explicitly rejected by educators. Indicators of rejection included follow-up requests expressing, "You are wrong," "Here is a mistake," "No, this is not what I described," etc. Additionally, if the AI responded with a question and educators chose not to answer, this was also counted as a rejection. Conversely, we recorded explicitly accepted responses where educators expressed gratitude or compliments in their follow-up requests or provided answers to AI initiated questions. Given the variability in personal communication styles among educators, without explicit expressions, it was challenging to gauge educators' attitudes towards the rest of the responses. Finally, we compared the evaluation results of rejected and accepted responses against the mean of the rest of the sample to add a layer of validation to our quality measures for AI response quality. This step ensured that our assessment criteria and LLM tools effectively reflected and captured the educators' views on the quality and applicability of AI responses within educational settings.

## 3.3 Alignment Analysis

### *3.3.1 Correlation Analysis*

To explore educators' needs for AI support concerning pedagogical aspects and instructional practices derived from their unique educational contexts, we conducted a correlation analysis at the conversational level. This choice—favoring conversational over request-level analysis—stems from the observation that educators may modify AI responses, leading to additional requests on pedagogy or practice that do not necessarily pertain to further implementation support. By analyzing at the conversation level, we ensure that the appearance of a label truly reflects an educator's spontaneous request for instructional preparation. We applied statistical correlation techniques to identify and quantify the relationships between pedagogical approaches and the educational contexts by assessing the strength and nature of these correlations.

We used Pearson's correlation coefficients to measure the degree of concurrent appearance between the specific instructional practice requests and contexts. Since we have conducted multiple iterations of coding inspection, refinement, and verification, this approach ensures that our analysis accurately reflects the nuanced ways in which instructional practices are implemented in different educational settings at the conversational level. Our findings not only provide a clear view of the alignment between instructional practices and educational needs but also underscore the potential for AI to enhance and support these practices across varied learning environments. This methodology is crucial for the development of AI tools that are both responsive and effective, meeting the dynamic needs of educators and learners.

### *3.3.2 Indicators for Context Revealing*

From our context analysis and labeling, we derived labels for educational contexts and instructional practices at both the conversation level and the request level nested within conversations. Combining these two types of labels allowed us to examine the adaptability of AI responses not only by comparing average response quality among conversations under given contexts for certain instructional practices but also along the conversation flow, particularly before and after the revealing of educational contexts.

During our study of educator-AI interactions, we noted variations in communication styles. Some educators provide comprehensive information in their initial message and focus on editing specific parts of the response in subsequent interactions. Due to AI memory within a conversation, the educational contexts revealed in the first message set the overarching background for the entire conversation, causing AI responses to be contextually adaptive from the outset. Conversely, some educators guided the AI conversations step-by-step, introducing or implying educational contexts during the conversation. In these cases, AI responses prior to detecting the context may not be contextually adapted.

To address these dynamics, we introduced two indicator variables for each conversation to denote requests made before and after the revealing of educational contexts and instructional practices, respectively. For requests before revealing context or practices, the corresponding indicators, "No Context" or "No Practice," are set to zero. This nuanced approach helps to precisely measure the contextual adaptability of AI responses, enhancing our understanding of AI's utility in diverse educational interactions.

# 4. RESULTS

## 4.1 RQ1. Conceptualization of AI in Educator Support

During the inductive coding for educational contexts, our analysis identified specific student needs and classroom settings that shape the educational contexts educators bring into AI interactions, as summarized in Table 1. These educational contexts encompass 41.56% of the conversations about mathematics education, indicating significant areas of focus for educators utilizing AI support.

**Table 1. Educational Contexts Represents in Educator-AI Conversations**

| Revealed Educational Contexts | Percentage |
|---|---|
| Mixed Ability Learners | 14.58 |
| Reluctant Participants or Engagement | 5.06 |
| Below Grade Level | 4.89 |
| Classroom Management and Climate | 4.17 |
| Advanced or Gifted Learners | 3.64 |
| Special Education (SpEd) | 3.19 |
| English Language Learners (ELL) | 3.03 |
| Behavioral and Emotional Support | 2.64 |
| Low-Tech Educational Environment | 0.36 |
| No Revealed Context | 58.44 |

The data reveals that educators frequently address the needs of mixed ability learners, which is more common than addressing the needs of single learning skills, such as advanced or gifted learners. Beyond the academic needs, educators also encounter challenges related to learners' attitudes, such as reluctance to participate in class activities or concerns about fostering a positive learning environment. The low frequency of discussions about low-tech educational environments may indicate a potential underutilization of AI, either because educators perceive it as ineffective in low-tech settings or due to a lack of access to necessary technological tools among educators and students.

Figure 3 shows the landscape of instructional practices observed in an AI-powered instruction preparation environment. The instructional practices are nested in pedagogical aspects aligned with effective instructional dimensions as identified. This figure illustrates where educators are focusing their AI conversations and highlights how AI can support and enhance their pedagogical practices. The analysis shows that educators are seeking AI solutions that facilitate differentiation, assessment, real-world applications, high-level critical thinking, and student engagement. These practices suggest a move beyond mere automation of rote tasks towards a more integrated approach where AI actively enriches teaching and learning experiences.

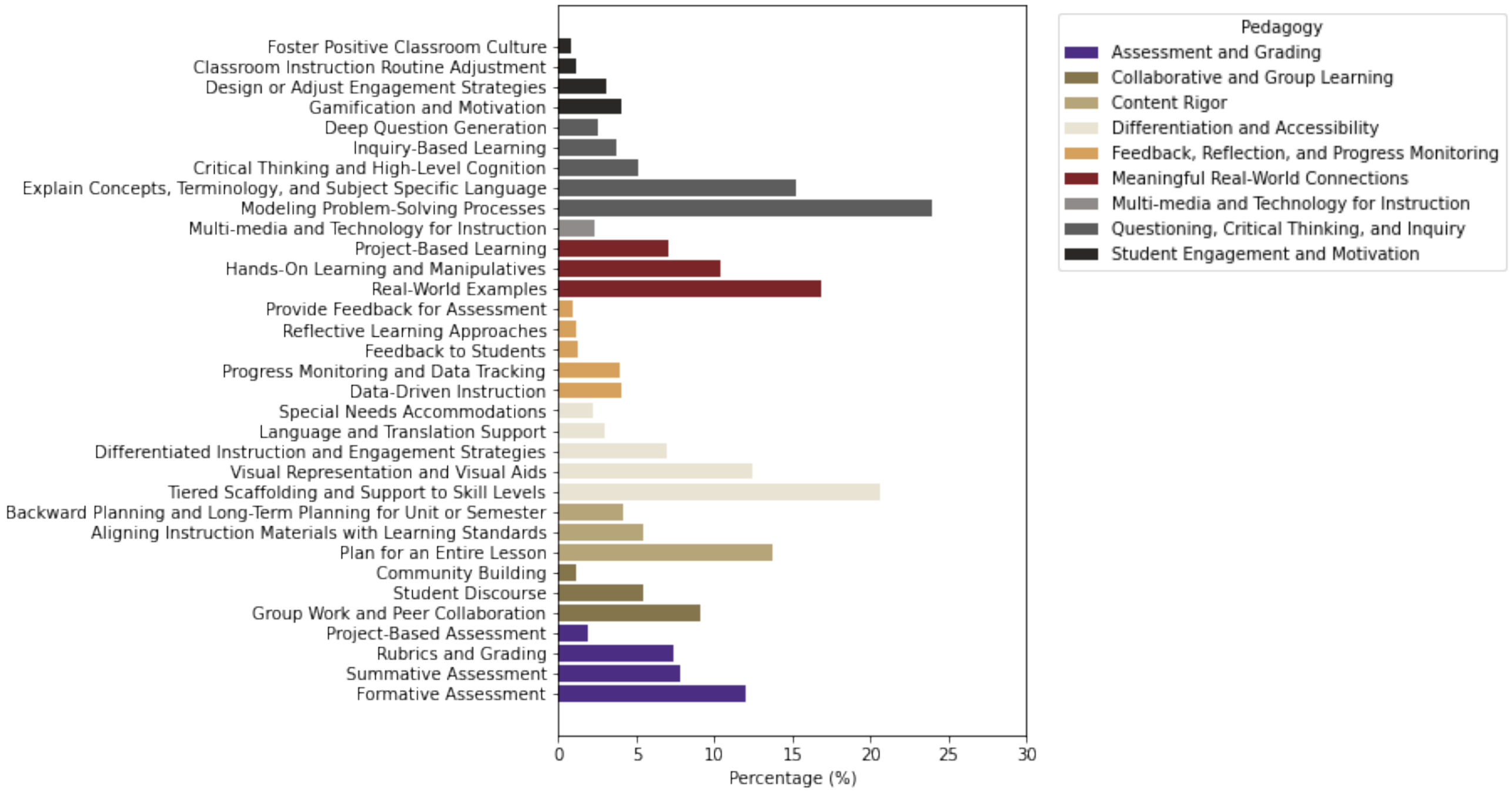

**Figure 3. Percentage of Conversations Inquiring About Instructional Practices Grouped by Pedagogy**

One of the main practices within "Differentiation and Accessibility" is tiered scaffolding and support (20.6%) which illustrates that teachers are seeking AI assistance in tailoring lessons to varied student abilities. The focus on visual aids (12.5%) suggests that tools like AI-powered visual generators could enhance multimodal learning. Issues like language barriers and special needs accommodations, though less frequently discussed (2.9% and 2.2% respectively), underscore the importance of accessible educational tools that adapt text and provide assistive features.

Within "Questioning, Critical Thinking, and Inquiry", educators emphasize the need for AI to assist in explain[ing] concepts, terminology and subject-specific language and in modeling problem-solving processes (23.9% and 15.2% respectively). This implies a role for AI in generating inquiry-based learning experiences and supporting educators to guide students in exploring complex ideas. Educators also emphasize the practices building "Meaningful Real-World Connections" through real-world problems and project-based learning (16.8% and 7.0%). This indicates demands for AI not only to generate the learning materials in students' real-world context but also to generate implementation plans for educators to apply the materials in meaningful ways.

Within "Assessment and Grading", create formative and summative assessments (12.0% and 7.8%) suggests a use for AI tools to create assessments. AI support on aligning existing materials with specify learning standards or generate new instructional materials (5.5% and 13.7%) improve the "Content Rigor" of instructional materials, which addresses a major concern for mathematics instruction [39]. Educators seek AI that can help organize and enhance lesson planning to ensure that educational content meets learning standards and is tailored to student needs. "Collaborative and Group Learning" is discussed with a focus on creating procedures for group work and peer collaboration while promote student discourse opportunities (9.1% and 5.4%). For "Feedback, Reflection, and Progress Monitoring", AI are leveraged for their importance in providing data-driven insights into student progress (4.1% and 4.0%) to inform instructional adjustment.

Overall, the results demonstrate the varied requests of teachers for AI in meeting complex educational demands and underscores the broad potential for AI to amplify educational outcomes by adapting and enhancing various instructional practices.

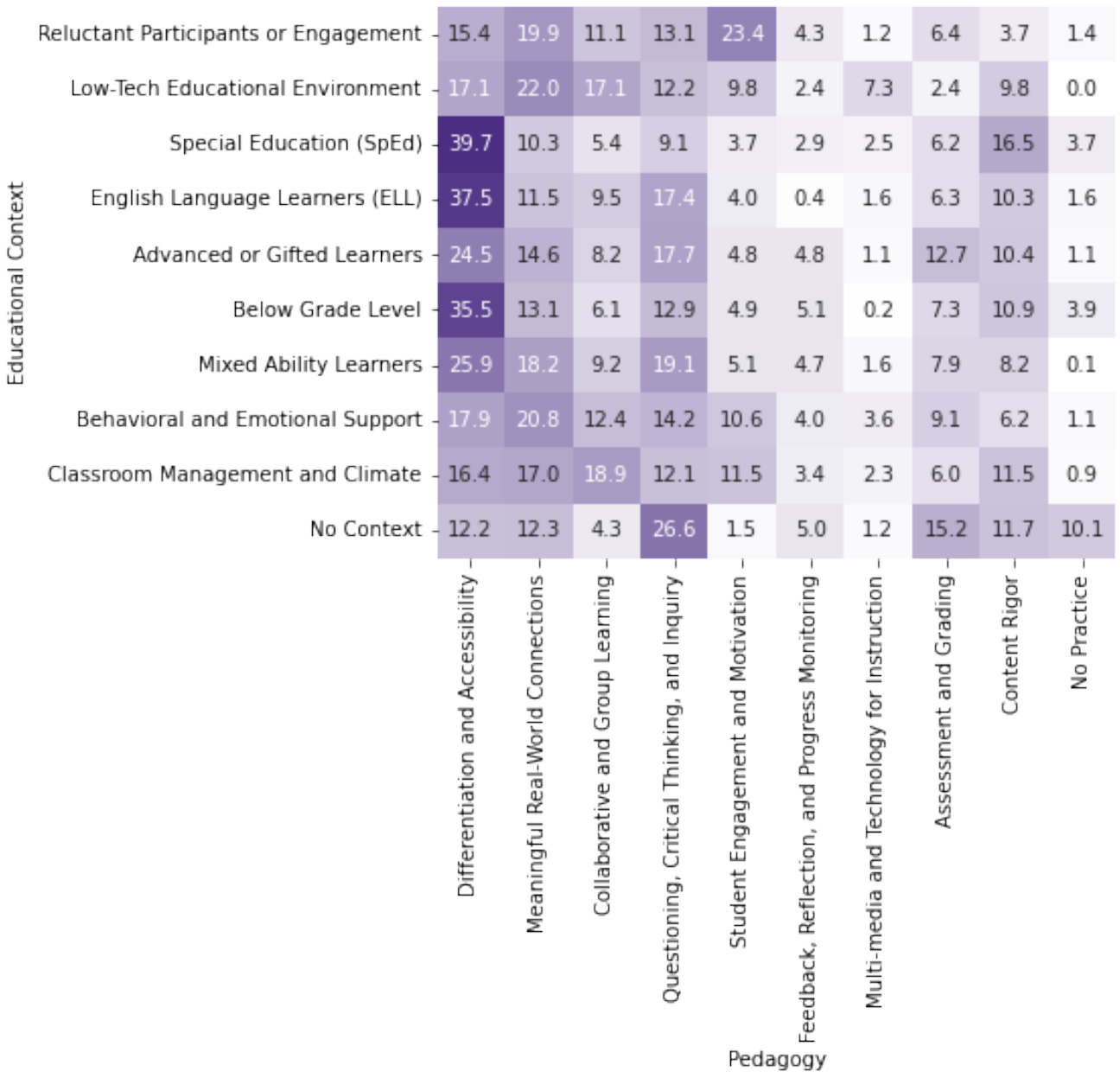

**Figure 4. Frequency of Instructional Practices Across Educational Contexts**

To determine how educational contexts influence educators' requests for instructional practices, we also performed correlation analysis between specific pedagogical focuses and educational contexts. Our results in figure 4 shows AI's nuanced application stems directly from the distinct needs arising within the educational settings where educators operate. For instance, environments catering to “Special Education (SpEd)” and “English Language Learners (ELL)” prominently emphasize “Differentiation and Accessibility”. This highlights a strong focus on personalized pedagogical techniques tailored to distinct learning needs and capabilities. Likewise, “Meaningful Real-World Connections” and “Collaborative and Group Learning” are highly prioritized in contexts that emphasize engagement, such as “Behavioral and Emotional Support” and “Classroom Management and Climate”. This suggests that educators value the integration of real-world applicability and cooperative learning strategies as effective methods for engaging students and managing complex classroom dynamics.

The educational context also significantly influences the types of information educators seek. For example, instead of widely requesting actionable implementation strategies, in “Low-Tech Educational Environment”, there is a marked preference for AI support in accessing real-world examples, lesson plans, and instruction for multimedia integration, indicating potential gaps in information accessibility within these settings.

A closer examination of instructional practices within these contexts reveals that support requests often align closely with established pedagogical approaches. In the realm of “Differentiation and Accessibility”, for instance, educators in SpEd settings frequently request accommodations that fit the descriptions of students’ Individualized Education Programs (IEPs), while ELL educators focus on language and translation support, reflecting the specific requirements of their respective contexts. Deeper analysis shows that language and translation support also differ; educators proficient in their students’ home languages often seek pedagogical strategies for vocabulary development in mathematics subject-specific languages, whereas educators who do not speak the home language may need frequent translations, underscoring the critical importance of translation accuracy of AI responses. The utilization of AI to generate “Tiered Scaffolding and Support to Skill Levels” within the same activity highlights its role in addressing the various needs present in educational settings with “Mixed Ability Learners”. “Visual Representation and Visual Aids” as multimodal learning tools are also significantly used in these contexts to accommodate various learning styles and preferences.

These insights align with previous findings about the role of AI in enhancing educational practices, especially in how AI tools can be strategically deployed to support differentiated instruction and real-world applications effectively. AI’s capability to offer tailored recommendations is particularly transformative in settings like special education and mixed-ability classrooms, where customized activities and generating visual aids can significantly enhance student engagement and comprehension. Additionally, the strategic use of AI to foster questioning, critical thinking, and inquiry in classrooms for advanced or gifted learners showcases its potential to challenge students cognitively and support educators in implementing inquiry-based learning models that promote higher-order thinking skills.

In conversations lacking a specific educational context, there is a notable emphasis on problem-solving and assessment. Educators frequently request AI assistance in developing problem-solving processes and creating assessment items that incorporate real-world applications, highlighting the widespread application of AI in enhancing instruction.

## 4.2 RQ2. AI Response Quality Measures

To validate our quality measures and ensure they accurately reflect educators’ views on AI response quality and applicability, we conducted a comparative analysis of rejected and accepted responses, assessing their accuracy, relevancy, and usefulness. Table 2 illustrates that responses deemed low in these dimensions were typically rejected, whereas those rated highly were accepted, closely aligning with the overall average ratings.

**Table 2. Average AI Response Quality by Follow-up Dynamics**

| | **Response Accuracy (0-3)** | **Response Relevancy (0-3)** | **Response Usefulness (0-3)** |
|---|---|---|---|
| Overall Average (N=8576) | 2.87 | 2.84 | 2.79 |
| Accepted (N=2230) | 2.85 | 2.93 | 2.87 |
| Rejected (N=367) | 2.55 | 2.44 | 2.22 |

This data corroborates our hypothesis: educators prefer AI responses that are not only accurate but also relevant and useful, rejecting those that fail to meet these criteria. The analysis substantiates the effectiveness of the established quality measures and affirms the essential roles of accuracy, relevancy, and usefulness in determining the applicability of AI tools in educational settings.

We then assessed AI response quality by analyzing it under four different conditions: without any educator stated context or practice, with only context revealed, with only practice revealed, and with both context and practice revealed. Our analysis at the conversation level indicated that conversations associated with revealed context or practice were more relevant and useful than those where no such information was disclosed. Conversations where both context and practice were revealed demonstrated the highest levels of relevancy and usefulness. However, the accuracy of AI responses did not exhibit a similar pattern; even without revealed context or practice, AI was able to maintain a high level of accuracy. This suggests that while the specificity of educator inputs (context or practice information) enhances the relevancy and usefulness of AI responses, generic model training is sufficient for maintaining accuracy [46].

To address potential concerns that AI response quality is dependent on the precision of educators’ input, as educators who are more skillful with AI can provide more specific requests containing context and practice information, we refined our approach by shifting from conversation-level to request-level analysis. We categorized requests based on their occurrence before and after the revelation of context or practice information. This refined analysis corroborated our earlier findings. This differentiated analysis underlines how the revelation of context or practice can influence the relevancy and usefulness of AI responses. Our findings align with our measurement criteria, which are crucial for the next steps in our analysis, focusing on the adaptability of AI responses to meet the various needs of educators across various contexts.

## 4.3 RQ3. Adaptability of AI Responses Across Educational Contexts

To assess the adaptability of AI supporting practices within different educational settings, we analyzed its performance across grouped conversations, categorized by specific educational contexts. We observed that AI generally maintained a high quality of response across all measured attributes—accuracy, relevance, and usefulness—especially notable in cases involving Content Rigor, see Figure 5 for a summary. AI’s adaptability shone in environments dealing with mixed-ability learners and specific classroom management needs, where it provided responses that were not only contextually tailored but also actionable and supportive.

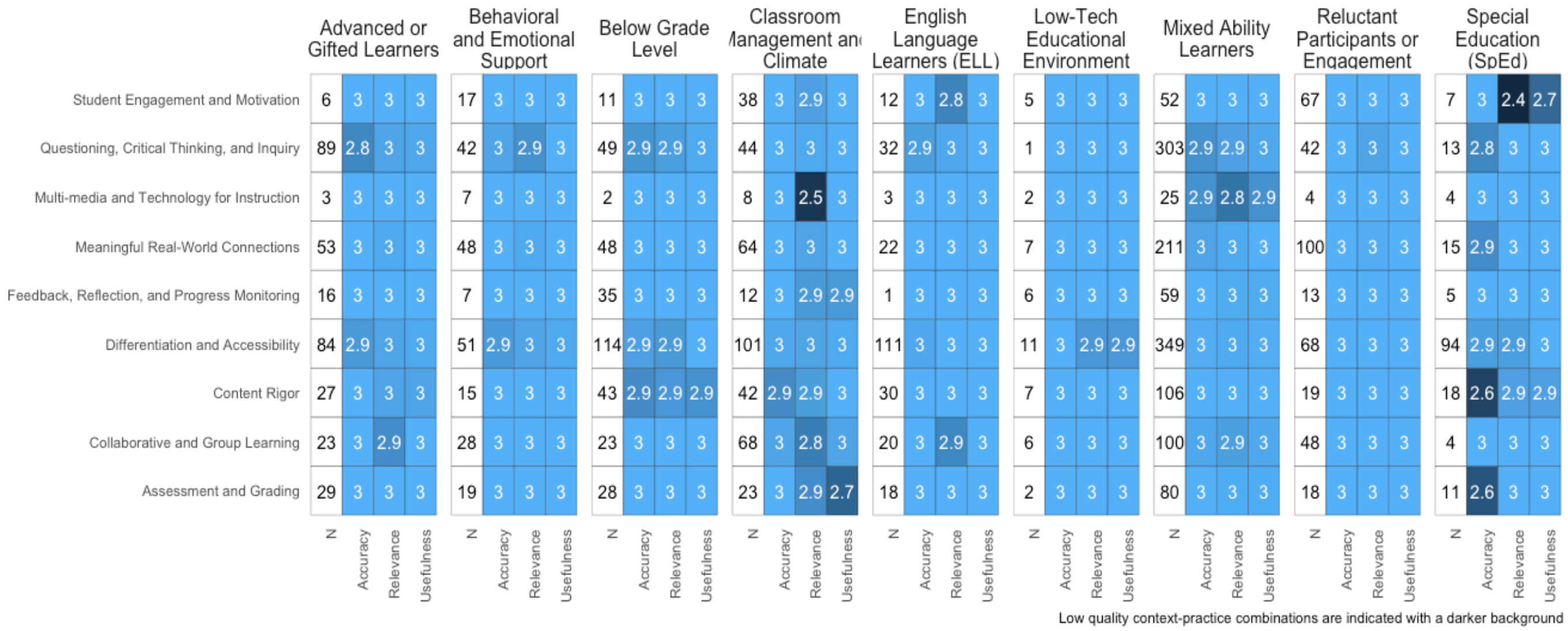

**Figure 5. LLM Response Quality By Educational Context and Pedagogical Request**

However, the adaptability of AI responses was not uniformly effective across all contexts. In situations requiring accommodating special needs for advanced learners or addressing behavioral and emotional needs, AI's responses sometimes lacked accuracy, leading to potential misunderstandings that could adversely affect the educational process. This was also evident in support for ELLs, where AI struggled to generate deep questions and tailor responses to the unique needs of these students and the educators, sometimes resulting in unproductive communication.

The integration of specific educational contexts into AI's decision-making process introduced complexities that interrupt its chain of thoughts, causing slightly degraded response quality. For instance, when educators sought AI assistance for "Visual Representation and Visual Aids", a specified educational context may reduce the quality of output slightly. Besides, when enhancing "Critical Thinking and High-Level Cognition" within classrooms with reluctant participants, the AI's ability to adapt was less effective, suggesting a misalignment between AI's inference capabilities and the nuanced demands of varied educational environments.

In addition, AI responses occasionally offer support that may extend beyond the conventional resources typically available in K-12 mathematics educators, which can diminish the relevance and utility. For example, when educators seek assistance with classroom management regarding integrating multimedia or technological support, AI might suggest the educators build an interactive board to engage students. Such recommendations, although innovative, often do not align with the standard responsibilities of a K-12 mathematics educator and may not be feasible to implement due to resource limitations or curriculum constraints.

Furthermore, the adaptability of AI can also be underscored by its self-learning capabilities, which were evident when it proactively solicited additional context or instructional details lacking in the initial conversations. Of the 1,722 conversations that involved revealed practices, 410 were enhanced by questions generated by the AI itself. Similarly, out of 623 conversations where educational contexts were revealed, 93 were initiated by AI's inquiries.

The strategy of AI initiating conversations with clarifying questions about context and practice carries potential drawbacks. Our observations indicate that when AI starts a dialogue with such questions, it can negatively impact user engagement—evidenced by 50 instances where educators chose to discontinue the conversation. This suggests a delicate balance in designing AI tools that are proactive yet user-friendly, ensuring they enhance usability without overwhelming educators.

# 5. DISCUSSION

Our findings for RQ1 suggest that educators envision AI not just as a tool for automation but as an intelligent collaborator that deeply enriches instructional practices. The data show that educators most frequently seek AI support for tasks like adaptive differentiation, facilitation of critical thinking, real-world learning applications, personalized assessment, collaborative engagement, and instructional planning. This indicates a demand for AI systems that transcend basic content delivery to offer dynamic, responsive, and context-sensitive support. Across different educational contexts, the specific needs and settings influence not only the practical support educators seek but also the format of that support, such as supplemental materials, assessment assistance, or actionable pedagogical strategies.

RQ2 findings validate accuracy, relevancy, and usefulness as AI applied metrics for evaluating AI response quality. The correlation between these metrics and educators' acceptance or rejection of responses supports our hypothesis that higher-quality responses are more likely to be accepted, underscoring the need to enhance AI capabilities to consistently meet high standards. Results also indicate that AI responses containing contextual or practice-related information are typically more relevant and useful, suggesting that enriching AI training with a broader understanding of educational contexts could significantly improve its performance and reliability.

RQ3 findings indicate the adaptability of AI responses to specific educational settings, like differentiating instructions for mixed ability learners, engagement strategies for classroom management needs, and enhancing cognitive demand for advanced learners. Our findings underscore the capacity of AI tools to integrate contextual information, essential for targeted instructional support. However, challenges persist in AI's ability to connect seemingly irrelevant context to specific suggested practices, such as promoting critical thinking strategies for teachers concerned about reluctant participation. Moreover, AI underperforms when contextual information adds additional complexity to already challenging tasks for AI like visualization. Grounding AI with realistic suggestions is also important for its effective support to K-12 education. The limitations in current AI tools to handle specific or less common educational scenarios indicate a direction for further development.

Furthermore, the capability of AI to proactively request additional contextual or instructional information marks a significant advancement in AI's adaptive capabilities. Yet, the associated risk of reduced usability, as some educators disengage when AI initiates conversations requiring extensive information input, requires a careful balance between AI's proactive behavior and maintaining user engagement for the continued development and acceptance of AI tools in education.

# 6. LIMITATION AND FUTURE STUDY

The current study, while shedding light on the application of AI in educational environments, encounters several limitations:

- **Exclusion of Interdisciplinary Conversations**: The dataset primarily focused on mathematics conversations, excluding interdisciplinary dialogues that incorporate math. This could limit our understanding of AI's utility across broader curriculum integrations and may not fully represent the diversity of educational contexts and instructional practices.
- **Incomplete Educator Information**: The absence of detailed information about educators, including specifics such as grade levels or years of experience, requires this analysis to rely on contextually revealed information in teacher-AI conversations. Further analysis that links teacher users to administrative data about their teaching context will allow a more robust validation of these findings.
- **Voluntary Participation and Geographic Imbalance**: Teachers in the study were early technology adopters primarily located on the US West Coast and may not be broadly representative of K-12 mathematics teachers.

To enhance the understanding of AI's role in educational contexts, future research should:

- **Integrate Detailed Educator and School Data**: Incorporate comprehensive information about educators' professional backgrounds and the schools or districts they belong to. This detailed context will allow for more nuanced analyses of how various educational environments influence the effectiveness of AI tools, offering insights that are finely tuned to specific local policies, resources, and needs.
- **Deepen Analysis of Instructional Practices**: Focus on a granular examination of specific request and response content regarding instructional practices within different educational settings. Such detailed text analysis could explore: the types of implementation strategies offered by the AI and the which mathematics concepts teachers requested explanations for?
- **Develop Anticipatory AI Tools**: Explore the creation of AI tools that utilize identified connections between practices and contexts to proactively meet educators' needs. These anticipatory tools could predict instructional challenges and offer preemptive support, potentially enhancing educational outcomes and optimizing teaching processes.

While we are unable to share the data underlying this study, our analytical code is publicly available at github.

# 7. ACKNOWLEDGMENTS

This work is supported by the Institute of Education Sciences of the U.S. Department of Education, through Grant R305C240012 and by several awards from the National Science Foundation (NSF #2043613, 2300291, 2405110) to the University of Washington, and a NSF SBIR/STTR award to Hensun Innovation LLC (#2423365). The opinions expressed are those of the authors and do not represent views of the funders.